\documentclass[twocolumn,epjc3]{svjour3}  

\usepackage{enumitem}
\usepackage{graphicx}
\usepackage{subfigure}
\usepackage{dcolumn}
\usepackage{array,multirow}
\usepackage{bm}
\usepackage{amsmath}
\usepackage{epstopdf}
\usepackage{amsfonts}
\usepackage{amssymb}
\usepackage{epsfig}
\usepackage{tabularx}
\usepackage{float}
\usepackage{color}
\usepackage[colorlinks=blue,citecolor=blue,urlcolor=blue,breaklinks]{hyperref}
\smartqed  
\RequirePackage{graphicx}
\usepackage{cite}

\providecommand{\U}[1]{\protect\rule{.1in}{.1in}}

\newcommand{\baa}{\begin{align}}
\newcommand{\eaa}{\end{align}}

\newcommand{\be}{\begin{equation}}
\newcommand{\ee}{\end{equation}}
\newcommand{\bea}{\begin{eqnarray}}
\newcommand{\ena}{\end{eqnarray}}

\journalname{XXX
}

\begin{document}

\title{Minimal Geometric Deformation in a Reissner-Nordstr\"om background}

\author{
\'Angel Rinc\'on \thanksref{e1,addr1} 
\and
Luciano Gabbanelli \thanksref{e2,addr2}
\and
Ernesto Contreras \thanksref{e3,addr3} 
\and
Francisco Tello-Ortiz \thanksref{e4,addr4}
}


\thankstext{e1}{angel.rincon@pucv.cl}
\thankstext{e2}{gabbanelli@icc.ub.edu}

\thankstext{e3}{econtreras@yachaytech.edu.ec}

\thankstext{e4}{francisco.tello@ua.cl}


\institute{
Instituto de F{\'i}sica, Pontificia Universidad Cat\'olica de Valpara{\'i}so, Avenida Brasil 2950, Casilla 4059, Valpara{\'i}so, Chile \label{addr1}
\and \
Dept. de F´ısica Qu`antica i Astrof´ısica, Institut de Ci`encies del Cosmos (ICCUB),
Universitat de Barcelona, Mart\'i i Franqu`es 1, 08028 Barcelona, Spain. \label{addr2} 
\and \ 
School of Physical Sciences \& Nanotechnology, Yachay Tech University, 100119 Urcuqu\'i, Ecuador\label{addr3}
\and \
 Departamento de F\'isica, Facultad de ciencias b\'asicas,
Universidad de Antofagasta, Casilla 170, Antofagasta, Chile.\label{addr4}}

\date{Received: date / Revised version: date}

\maketitle

\begin{abstract}
This article is devoted to the study of new exact analytical solutions in the background of Reissner-Nordstr\"{o}m space-time by using gravitational decoupling via minimal geometric deformation approach. To do so, we impose the most general equation of state, relating the components of the $\theta$-sector in order to obtain the new material contributions and the decoupler function $f(r)$. Besides, we obtain the bounds on the free parameters of the extended solution to avoid new singularities. Furthermore, we show the finitude of all thermodynamic parameters of the solution such as the effective density $\tilde{\rho}$, radial $\tilde{p}_{r}$ and tangential $\tilde{p}_{t}$ pressure for different values of parameter $\alpha$ and the total electric charge $Q$. Finally, the behavior of some scalar invariants, namely the Ricci $R$ and Kretshmann $R_{\mu\nu\omega\epsilon}R^{\mu\nu\omega\epsilon}$ scalars are analyzed. It is also remarkable that, after an appropriate limit, the deformed Schwarzschild black hole solution always can be recovered.
\end{abstract}

%
%
%

\maketitle

\section{Introduction}

Black hole idea has a long history. At first, Newton's universal gravitation theory was used to investigate the existence of \emph{dark stars}. Nonetheless, the starting point to corroborate the existence and understanding of the behavior of these peculiar structures dates back to 1915, when Albert Einstein made known his famous general theory of relativity (GR). Shortly after the publication of GR, K. Schwarzschild \cite{r1} was the first to report a solution to Einstein's field equations. This solution describes a spherically symmetrical and static object without electric charge (it is the only known vacuum solution of Einstein's field equations), a super-dense region of space-time that exhibits a strong gravitational field where nothing can escape (matter, not even the light).
The formation of these structures in the Universe is due to the gravitational collapse of massive stars (20 times more than the mass of the Sun). These black holes are called stellar black holes, while those formed by the collapse of stars much more massive than $20 M_{\odot}$ are called supermassive black holes ($10^{6}$ times the mass of the sun). For a black hole to be created, the collapsed star will shrink down to an infinitely dense point called a singularity. This singularity is a region of the space-time where the laws of physics are no longer valid. Due to the cosmic censorship hypothesis, all singularities should be covered by a  line called the event horizon. It means that naked singularities are forbidden. Notably, the well known Schwarzschild black hole has an essential singularity at $r=0$ protected by an event horizon at $r=2M$ (where $M$ is the total mass). Furthermore, these impressive structures are characterized by three conserved charges: I) the mass $M$, II) the electric charge $Q$ and III) the angular momentum $J$. 
Although the above is also true, the Schwarzschild solution is described only by the mass $M$ parameter, the Reissner-Nordstr\"{o}m solution \cite{r2,r3} is characterized by mass $M$ and electric charge $Q$ whereas the Kerr space-time \cite{r4} is painted by mass $M$ and angular momentum $J$ charges. Moreover, the most general solution of this type is the Kerr-Newman space-time \cite{r5} characterized by mass $M$, electric charge $Q$, and angular momentum $J$. The existence of these conserved charges is supported by the non-hair conjecture \cite{r6}, which states that these solutions should not carry any other charges. Nonetheless, internal gauge symmetries and extra fields could introduce new conserved charges, such as soft quantum hair \cite{r7}.

Form the theoretical point of view, these exciting objects have been extensively studied and classified \cite{r8}. On the other hand, the existence of these intriguing objects was observationally corroborated in 2016 by  LIGO, Virgo, and GEO600 collaborations. This announ\-ced was the first detection of gravitational waves produced by the fusion of two black holes \cite{r9,r10,r11}. This observational evidence called GW150914 was an unprecedented event that gave further support to Einstein's theory. Notwithstanding, theoretically speaking, there are many questions to answer about these fascinating celestial bodies. Despite its mathematical beauty, handling problems of physical relevance in GR is usually a formidable task. Since it is a highly non-linear theory, the principle of superposition valid in linear differential equations does not apply here and finding exact solutions has always been a challenge. 
Even more, black hole solutions have been analyzed in several dimensions. Besides, they have been studied classically and after that, under a quantum point of view. The solutions mentioned above (Schwarzschild,  Reissner-Nordstr\"{o}m and Kerr-Newman) include classical effects only, but new and exciting effects can be included when we relax some of the usual assumptions made in GR. For instance, the well-known RG-improvement technique also incorporates quantum features into classical solutions 
\cite{
Bonanno:2000ep,
Bonanno:2006eu,
Bonanno:2016ako,
Reuter:2000nt,
Reuter:2003ca
}. 
Following similar ideas, the so-called scale-dependent scenario (which is Asymptotic safety inspired) accounts quantum effects via the running of the gravitational coupling 
\cite{
Koch:2016uso,
Rincon:2017ypd,
Rincon:2017goj,
Rincon:2017ayr,
Contreras:2017eza,
Rincon:2018sgd,
Hernandez-Arboleda:2018qdo,
Contreras:2018dhs,
Rincon:2018dsq,
Contreras:2018gct,
Canales:2018tbn,
Rincon:2019cix,
Rincon:2019zxk,
Contreras:2019fwu,
Fathi:2019jid,
Contreras:2019cmf,
Contreras:2018gpl,
Contreras:2018swc
}. 
Both methods modify the classical BH solutions assuming that the coupling parameters are not constants anymore. This assumption can also be interpreted as an anisotropic energy-momentum coming from the quantum sector. Thus, anisotropic solutions could appear when quantum features are present. The study of anisotropic solutions in black holes is two folds: firstly, to get insights about the underlying physics in anisotropic black hole solutions, and second, to establish, if this exists, a connection between anisotropies and physics beyond Einstein gravity.

In order to generate anisotropic solutions, a new and elegant method that allows us to obtain new exact solutions starting from a known one has received considerable attention recently \cite{r56,r57}. The so-called gravitational decoupling through Minimal Geometric Deformation (MGD henceforth) approach, was developed to deform Schwarzschild space-time \cite{r58,r59} in the Randall-Sundrum brane-world \cite{r60,r61}. Basically, this grasp works by extending simple solutions into more complex domains, which serve up to explore new insights in diverse areas. The full history of how this methodology was developed and how it works can be found in the following references \cite{r62,r63,r64,r65,r66,r67,r68}.

In recent years, there has been a growing interest in using this machinery to explore the behavior of collapsed structures, such as neutron stars and black holes in the presence of anisotropic matter distributions. Particularly, models representing  perfect fluid spheres without electric charge/with electric charge have been extended to anisotropic domains \cite{r69,r70,r71,r72,r73,r74,r75,r76,r77,r78,r79,r96}. Besides, black hole solutions have also been addressed within the MGD arena; specifically, the Schwarzschild space-time \cite{r80}, BTZ manifold \cite{r81} and AdS geometry \cite{r82} have been worked. Also the inverse problem was addressed in $3+1$ dimensions \cite{r83} and $2+1$ dimensions including cosmological constant \cite{r84}.  Besides, a cloud of string \cite{r85} and Klein-Gordon scalar fields as an extra matter content \cite{r86} were treated. Although the method was developed for spherically symmetric geometries, it was spread out to be used in isotropic coordinates \cite{r87}. Moreover, the existence of exotic structures such as ultra-compact \- Schwarzschild star, or gravastar \cite{r88} was investigated.

As well, in a broader context, the extension of the method, including geometric deformations on both metric potentials was reported at \cite{r89,r90}; it was called the extended-MGD scheme. On the other hand, given the abundance of modified gravity theories and treatments of the gravitational interaction in the regime of extra dimensions, the extension of the method in these scenarios is completely natural. As far as this is concerned, neutron stars have been studied very recently considering extra dimensions \cite{r91} and in the background of Pure Lovelock gravity \cite{r92}, f(R,$\mathcal{T}$) gravity theory \cite{r93}, cosmological scenario \cite{r94}, Rastall gravity \cite{r95} and Brane-worlds \cite{r97}.

Following the same spirit of these good antecedents in this work, we investigate how the contributions introduced by gravitational decoupling through MGD modify the material content and geometry of the well known Reissner-Nordstr\"{o}m space-time. What is more, we follow the same procedure as was done in \cite{r80}. Precisely, in order to obtain the decoupler function $f(r)$ the most general equation of state relating the components of the $\theta$-sector is imposed. It is worth mentioning that this scheme work preserves the critical point of the original solution \i.e, its essential singularities, inner and outer event horizons, however, introduces new ones. These new critical points could be interpreted as new event horizons or new charges (hair) coming from the anisotropic behavior inserted by the $\theta$-sector. Besides, we explore the behavior of the salient energy-momentum tensor via energy conditions. In our case, the energy-momentum tensor corresponds to an anisotropic charged one. The deportment of some scalar invariants such as Ricci and Kretschmann scalar are analyzed. As was pointed out earlier, the existence of new fundamental fields, which yields to hairy black hole solutions, in the background of Reissner-Nordstr\"{o}m space-time is precisely the focus under study in this paper.
So, the present manuscript is organized as follows: in Sec. \ref{SEC1} gravitational decoupling field equations and MGD procedure are presented, Sec. \ref{RNMGD} the $\theta$-sector in solved by imposing the most general equation of state, and the new solution is depicted analyzing the behavior of the main salient thermodynamic functions and some scalar invariants. Finally, Sec. \ref{conclu} summarizes and concludes the reported study. We adopt the most negative metric signature, $(+,-,-,-)$.

\section{Field equations and Minimal Geometric Deformation}\label{SEC1}

\subsection{Einstein field equations}
\label{s2}
In curvature coordinates, a spherically symmetric and static geometry is described by the following line element  
\begin{equation}
ds^{2}=
e^{\nu (r)}\,dt^{2}-e^{\lambda (r)}\,dr^{2}
-r^{2}\left( d\theta^{2}+\sin ^{2}\theta \,d\phi ^{2}\right)
\ ,
\label{metric}
\end{equation}
where the metric functions, namely $\nu =\nu (r)$ and $\lambda =\lambda (r)$ are purely radial functions. With this geometry in hand and Einstein field equations
\begin{align}\label{EinEqFull}
R_{\mu \nu} - \frac{1}{2}g_{\mu \nu} R = -\kappa \widetilde{T}_{\mu \nu},
\end{align}
one obtains the following set of equations
\begin{align}\label{ec1}
\kappa \widetilde\rho &= \frac{1}{r^2}-e^{-\lambda}\left(\frac{1}{r^2}-\frac{\lambda'}{r}\right)\,,
\\\label{ec2}
-\kappa \widetilde{p}_r &= \frac{1}{r^2}-e^{-\lambda}\left(\frac{1}{r^2}+\frac{\nu'}{r}\right)\,,
\\\label{ec3}
-\kappa \widetilde{p}_t &=-\frac{1}{4}e^{-\lambda}\,\left(2\,\nu''+\nu^{\prime 2}-\lambda^{\prime}\,\nu^{\prime}+2\,\frac{\nu'-\lambda'}{r}\right)\,.
\end{align}
In the above system of equations the quantities $\tilde{\rho}$, $\tilde{p}_{r}$ and $\tilde{p}_{t}$ are the thermodynamic functions that characterize the energy-momentum tensor $\tilde{T}_{\mu\nu}$ in Eq. (\ref{EinEqFull}). These quantities are referred as the energy-density, the radial, and the transverse pressure, respectively. The overall constant $\kappa\equiv 8\pi G/ c^2$ throughout the article will be equal to $8\pi$ (i.e., relativistic geometrized units are employed $G=c=1$). Moreover, for the sake of simplicity, we shall use $\kappa=1$ in explicit computations.

Please, note that the linear combination of Eqs. (\ref{ec1})-(\ref{ec3}) invokes the conservation law (Bianchi's identity) of the energy-momentum tensor, given by
\begin{equation}\label{ConsEqEffective}
\nabla^\nu\,\widetilde{T}_{\mu\nu}=0\,.
\end{equation}
We should remark that throughout this article we will separate 
the energy-momentum tensor as follows
\begin{align}\label{StressTensorEffective}
\widetilde{T}_{\mu \nu} \equiv M_{\mu \nu} + \alpha \theta_{\mu \nu},
\end{align}
where the first term $M_{\mu \nu}$ encodes a known source, which, 
is a solution of the Einstein field equations with metric $ds^{2}=e^{\xi}dt^{2}-\frac{1}{\mu}dr^{2}+r^{2}(d\theta^{2}+\sin^{2}d\phi^{2})$, defined according to
\begin{align}\label{StressEnergyTensorPF}
M_{\mu\nu}=(\rho+p_{t})\,u_\mu u_\nu
-p_{t}\,g_{\mu\nu}+(p_{r}-p_{t})s_{\mu}s_{\nu}\,,
\end{align}
being $u^{\mu}$ and $s_{\mu}$ normalized four-velocity fields with the properties
$u_{\mu}u^{\mu}=s_{\mu}s^{\mu}=-1$ and $u_{\mu}s^{\mu}=0$. The Eq. (\ref{StressEnergyTensorPF}) is representing the most general form of an anisotropic matter distribution. In this respect, the input source $M_{\mu \nu}$ could be in principle anything \i.e, isotropic (with or without electric charge), anisotropic, electrically charged only.  In our case, the known solution is taken to be the Reissner-Nordstr\"{o}m space-time. The following metric potentials characterize this manifold
\begin{equation}\label{RNS}
e^{\xi}=\mu=1-2\frac{M}{r}+\frac{Q^{2}}{r^{2}},    
\end{equation}
being $M$ the mass of the object and $Q$ the total electric charge. Consequently, Eq. (\ref{StressEnergyTensorPF}) becomes
\begin{equation}\label{electrotensor}
M_{\mu\nu} = \text{diag} \left(\frac{E^{2}}{8\pi},-\frac{E^{2}}{8\pi},\frac{E^{2}}{8\pi},\frac{E^{2}}{8\pi}\right).    
\end{equation}
As should be noted the pure electromagnetic energy-momentum tensor (\ref{electrotensor}) is anisotropic in nature because $p_{r}\neq p_{t}$. Of course $p_{r}=-p_{t}=E^{2}/8\pi$. Furthermore, the electric field $E$ established a privileged direction which breaks down the isotropy.  
Also, the second term $\theta_{\mu \nu}$ parametrizes any additional unknown source which is coupled to gravity via the dimensionless $\alpha$ parameter.

So, the conservation equation \eqref{ConsEqEffective} produces
\begin{equation}\label{CON}
\nabla^{\nu}\tilde{T}_{\mu\nu}=\nabla^{\nu}F_{\mu\nu}+\alpha\nabla^{\nu}\theta_{\mu\nu}=0,    
\end{equation}
where $F_{\mu\nu}$ is the well known skew-symmetric Faraday-Maxwell electromagnetic tensor defined as
$F_{\mu\nu}=\partial_{\mu}A_{\nu}-\partial_{\nu}A_{\mu}$ with $A_{\mu}=\left(A_{0},0,0,0\right)$ the four-vector potential (as we shall consider a static configuration then $A_{i}=0$). Moreover, the electromagnetic tensor $F_{\mu\nu}$ satisfies the covariant Maxwell's equations 
\begin{align}
\label{maxwell1}
&\partial_{\mu}\left[\sqrt{-g}F^{\nu\mu}\right]=4\pi\sqrt{-g} J^{\nu},
\\
\label{maxwell2}
&\partial_{\alpha}F_{\beta\sigma}+\partial_{\beta}F_{\sigma\alpha}+\partial_{\sigma}F_{\alpha\beta}=0,
\end{align}
where $J^{\nu}$ is the electromagnetic four-current vector defined as
\begin{equation}\label{fourdensitycurrent}
J^{\nu}=\sigma u^{\nu},   
\end{equation}
representing $\sigma=e^{\xi/2}J^{0}(r)$ the charge density.
Concretely (\ref{CON}) leads to
	\begin{align}\label{ConsEqExplicit}
\begin{split}	
   \frac{qq^{\prime}}{4\pi r^{4}}  +
	\alpha 
	& 
	\Bigg[ 
	(\theta_r{}^r)' + 
	\frac{\xi'}{2}(\theta_r{}^r-\theta_t{}^t) \ + 
	\\
	& \ \ \ \ \ \ \ \ \ \ \ \ \  \frac{2}{r}\left(\theta_r{}^r-\theta_\varphi{}^\varphi\right)
	\Bigg]
	=0 
	\,,
\end{split}	
	\end{align}
with $q(r) \equiv  E(r)r^{2}$. At this point, we remark that in the present case the only non-vanishing component of the electromagnetic tensor is the electric field $E(r)=F^{01}=-F^{10}$. So, by simple inspection of the field equations~(\ref{ec1})--(\ref{ec3}), we can identify an effective density and two effective pressures, the first one $\tilde{p}_r$ is the radial pressure, whereas the second $\tilde{p}_t$ is the tangential pressure:
\begin{align}
\tilde{\rho} &= \frac{E^{2}}{8\pi} +\alpha\,\theta_0^{\,0} \ , \label{tildero}
\\
\tilde{p}_{r} &= -\frac{E^{2}}{8\pi}-\alpha\,\theta_1^{\,1} \ , \label{tildepr}
\\
\tilde{p}_{t} &= \frac{E^{2}}{8\pi}-\alpha\,\theta_2^{\,2} \ . \label{tildept}
\end{align}
This clearly illustrates that the source $\theta_{\mu\nu}$ modifies the anisotropy
\begin{equation}
\label{anisotropy}
\Delta
\equiv
\tilde{p}_{t}-\tilde{p}_{r}
=
\frac{E^{2}}{4\pi} - \alpha (\theta_2^2 - \theta_1^1).
\end{equation}
As can be seen, the system of Eqs.~(\ref{ec1})-(\ref{ec3}) contains five unknown functions, namely, 
three physical variables, the density $\tilde{\rho}(r)$, the radial pressure $\tilde{p}_r(r)$ and the tangential pressure $\tilde{p}_t(r)$, and two geometric functions: the temporal metric function $\xi(r)$ and the radial metric function $\mu(r)$. Therefore these equations form an indefinite system. In the following subsection we will face it by employing gravitational decoupling via MGD grasp as was mentioned above.

\subsection{Gravitational decoupling by MGD}

At the end of the previous section we agreed on separate the components of the energy--momentum tensor in a well--known matter sector and an extra unknown source $\theta_{\mu\nu}$. The next step, consist in to introduce a geometrical deformation
which allows to decouple the equations associated to the complex source $\tilde{T}_{\mu\nu}$ in a set of Einstein's equations sourced by a well known matter sector with metric functions $\{\xi,\mu\}$ and another fulfilling like--Einstein's equation sourced by $\theta_{\mu\nu}$ with metric functions $\{g,f\}$. The main goal is to use the known sector as a seed to solve the system $\{g,f,\theta_{\mu\nu}\}$. Finally, the strategy is to combine the results to obtain a solution for Eq. (\ref{EinEqFull}). Of course, give the non--linearity of the Einstein equations, the above protocol looks like a naive strategy. However, in the framework of MGD, the separation can be done in static and spherically symmetric space--times. If we consider
the metric
\begin{eqnarray}
ds^{2}=e^{\xi}dt^{2}-\frac{1}{\mu}dr^{2}-r^{2}(d\theta^{2}+ \sin^{2}\theta d\phi^{2}),
\end{eqnarray}
as a solution of the Einstein field equations sourced by a well known matter content, the most general geometric deformation that can be proposed reads
\begin{eqnarray}\label{expectg}
\xi &\to& \nu=\xi+ \alpha g, \\
\mu&\to&e^{-\lambda}=\mu+\alpha f. 
\end{eqnarray}
In this work, we are interested in the particular case $g=0$, such that all the sectors have the same $g_{tt}$ component.

Now let us plug the  decomposition in Eq.~(\ref{expectg}) in the Einstein equations~(\ref{ec1})-(\ref{ec3}). The system, as stated before, is thus  separated in two sets: (i)  having the standard Einstein field equations for an anisotropic fluid ($\alpha = 0$) of density $\rho$, radial pressure $p_r$, tangential pressure $p_t$, temporal metric component $g_{00}=e^{\nu}$ and radial metric component $g_{11}=-\mu^{-1}$ given by
\begin{eqnarray}
\label{ec1pf}
&&
-\kappa\rho
=-\frac{1}{r^2}+\frac{\mu}{r^2}+\frac{\mu'}{r}\ ,
\\
&&
\label{ec2pf}
-\kappa
\left(- p_r \right)
=
-\frac 1{r^2}+\mu\left( \frac 1{r^2}+\frac{\nu'}r\right)\ ,
\\
&&
\label{ec3pf}
-\kappa
\strut\displaystyle
\left(- p_t \right)
=
\frac{\mu}{4}\left(2\nu''+\nu'^2+\frac{2\nu'}{r}\right)+\frac{\mu'}{4}\left(\nu'+\frac{2}{r}\right)
\ ,
\end{eqnarray}
with the conservation equation yielding
\begin{equation}
\label{conpf}
\frac{qq^{\prime}}{4\pi}=0,
\end{equation}
which is a linear combination of Eqs~(\ref{ec1pf})-(\ref{ec3pf}); and (ii) for the source $\theta_{\mu\nu}$, which reads
\begin{eqnarray}
\label{ec1d}
&&
-\kappa\,\theta_0^{\,0}
=
\strut\displaystyle\frac{f}{r^2}
+\frac{f'}{r}\ ,
\\
&&
\label{ec2d}
-\kappa
\strut\displaystyle
\,\theta_1^{\,1}
= f\left(\frac{1}{r^2}+\frac{\nu'}{r}\right)\ ,
\\
&&
\label{ec3d}
-\kappa
\strut\displaystyle\,\theta_2^{\,2}
=\frac{f}{4}\left(2\nu''+\nu'^2+2\frac{\nu'}{r}\right)+\frac{f'}{4}\left(\nu'+\frac{2}{r}\right)
\ .
\end{eqnarray}
The conservation equation  $\nabla_\nu\,\theta^{\mu\nu}=0$ explicitly reads
\begin{equation}
\label{con1d}
(\theta_1^{\,\,1})'-\strut\displaystyle\frac{\nu'}{2}(\theta_0^{\,\,0}-\theta_1^{\,\,1})-\frac{2}{r}(\theta_2^{\,\,2}-\theta_1^{\,\,1}) = 0
\ ,
\end{equation}
which is a linear combination of Eqs.~(\ref{ec1d})-(\ref{ec3d}). 
Under these conditions, there is no exchange of energy-momentum between the perfect fluid and the source $\theta_{\mu\nu}$; their interaction is purely gravitational. 
%

\section{New exact solution in Reissner-Nordstr\"om background}\label{RNMGD}

\subsection{General constraint}

In order to obtain an analytical solution, we need to determine the deformation function $f(r)$. To do that, we will assume a particular condition between the components of the additional anisotropy $\theta_\nu^\mu$. Following \cite{r80}, we will impose the constraint
\be
\theta_0^0 = a \theta_1^1 + b \theta_2^2,
\ee
with two arbitrary constant parameters $a,b$. Using the equations satisfied by the deformation function, we obtain an ordinary differential equation of first order for $f(r)$ of the form
\be
\frac{\mathrm{d}f}{\mathrm{d}r} = -\frac{\beta}{\alpha} f,
\ee
where the functions $\alpha \equiv \alpha(r), \beta \equiv \beta(r)$ are found to be
\begin{align}
\alpha  = \ & \frac{1}{4} b \xi' + \frac{b}{2 r}-\frac{1}{r},  \\
\beta  = \ & \frac{a}{r^2} \Bigl(1 + r \xi' \Bigl) + \frac{1}{2} b \xi'' + \frac{1}{4} b (\xi')^2 +
\frac{b}{2 r} \xi' -\frac{1}{r^2},
\end{align}
The equation above can be integrated directly and we obtain for the deformation function the expression
\begin{align}
\begin{split}
f(r) = & \left[1- \frac{2 M}{r} + \frac{Q^2}{r^2}\right]
\left[1 - 
\frac{BM}{r}
+
\frac{C Q^2}{r^2} 
\right]^{A} 
\times
\\
&
\left[
\frac{
1-
\left(\frac{1}{2} + \frac{1}{2}\bar{R} \right) \frac{B M}{r}
}{
1+
\left(-\frac{1}{2} + \frac{1}{2}\bar{R} \right) \frac{B M}{r}
}
\right]^E \left(\frac{r}{L}\right)^F ,
\end{split}
\end{align}
where we have defined the following parameters:
\begin{align}
A &= \frac{a (b-4)-b^2+b+4}{2 (b-2)},
\\
B &= \frac{b-4}{b-2},
\\
C &= -\frac{2}{b-2},
\\
\bar{R} &= \sqrt{1 - 4 \frac{C}{B^2} \left(\frac{Q}{M}\right)^2 } ,
\\
E &= \frac{b (-a+b-1)}{2 (b-2) \bar{R}},
\\
F &= -\frac{2 (a-1)}{b-2}.
\end{align}
Please, notice that we have a first order differential equation for $f$, which means that we only have an integration constant $L$.
It is remarkable that, after demands $Q \rightarrow 0$, we recover the general expression previously obtained in \cite{r80} for the Schwarzschild vacuum solution, namely
\begin{align}
f(r)=\left(1-\frac{2M}{r}\right)\left(\frac{L}{r-BM}\right)^{\frac{2(a-1)}{b-2}}   . 
\end{align}
From Eq. (\ref{expectg}), it is straightforward to show that the $g^{rr}$ component of the total solution can be written as
\begin{align}
\begin{split}
e^{-\lambda} = &\left[1- \frac{2 M}{r} + \frac{Q^2}{r^2}\right]\mathcal{G}(r) ,
\end{split}
\end{align}
where
\begin{align}
\begin{split}
\mathcal{G}(r)=&1+\alpha
\left[1 - 
\frac{BM}{r}
+
\frac{C Q^2}{r^2} 
\right]^{A} 
\times
\\
&
\left[
\frac{
1-
\left(\frac{1}{2} + \frac{1}{2}\bar{R} \right) \frac{B M}{r}
}{
1+
\left(-\frac{1}{2} + \frac{1}{2}\bar{R} \right) \frac{B M}{r}
}
\right]^E \left(\frac{r}{L}\right)^F.
\end{split}
\end{align}
Please, be careful with the adequate selection of the parameters $\{a,b\}$. Although we can naively take arbitrary values of them, it could be better to analyze the critical points to be aware of them.
It is noticeable that the horizons of the extended solution coincide with those of the Reissner--Nordstr\"{o}m background. However, in order to avoid the apparition of extra singularities, an analysis on the critical points of the auxiliary function, $\mathcal{G}$ is mandatory. To be more precise, we need to explore the conditions to ensure that $\mathcal{G}$ is positive and finite everywhere. For example, note that in the case $E>0$ a critical point appears when $1+(-\frac{1}{2}+\frac{1}{2}\bar{R})\frac{BM}{r}=0$, which leads to
\begin{eqnarray}
r_{c}=-\frac{1}{2} B M (\bar{R}-1).
\end{eqnarray}
In order to avoid such a critical point, we could impose $r_{c}<0$, or, more precisely
\begin{align}
-\frac{(b-4)M}{2 (b-2)}
\left(\sqrt{1-\frac{4 (b-2)^2 C Q^2}{(b-4)^2 M^2}}-1\right)<0,
\end{align}
which can be satisfied in the following cases
\begin{align}
b<2\ &\textnormal{and}\ C<0,\\
2<b<4\ &\textnormal{and}\ 0<C\leq \frac{(b-4)^2 M^2}{4 (b-2)^2 Q^2},\\
b> 4\ &\textnormal{and} \ C<0.
\end{align}
Furthermore, the above requirements must be complemented with the extra condition $A>0$, to avoid singularities in the cases that the term 
$1-\frac{BM}{r}+\frac{C Q^{2}}{r^{2}}$ has real roots. In this respect, $A>0$ leads to the following constraints
\begin{eqnarray}
b<2\ &\textnormal{and}&\ a>\frac{b^2-b-4}{b-4},
\label{b1}
\\
2<b<4\ &\textnormal{and}&\ a<\frac{b^2-b-4}{b-4},\\
b>4\ &\textnormal{and}&\ a>\frac{b^2-b-4}{b-4}.
\end{eqnarray}
It is worth mentioning that with above constraints, the condition $\mathcal{G}>0$ is automatically satisfied.

In the following subsections we will consider a few concrete examples, although we still can get insights about the underlying physics at this level. First, the geometric deformation (\ref{expectg}) is proportional to the usual RN solution which is common feature of this formalism. 
Second, the present solution is based on the classical anisotripic case, i.e. the density and pressure (radial and tangential) are different from zero. The corresponding effective parameters are attributes to the anisotropic effect induced again by the method. Finally, we observe an unexpected feature: the solution is absent of new singularities just for certain concrete values of the free parameters $\{a,b\}$.
\begin{figure}[ht!]
\centering
\includegraphics[width=0.5\textwidth]{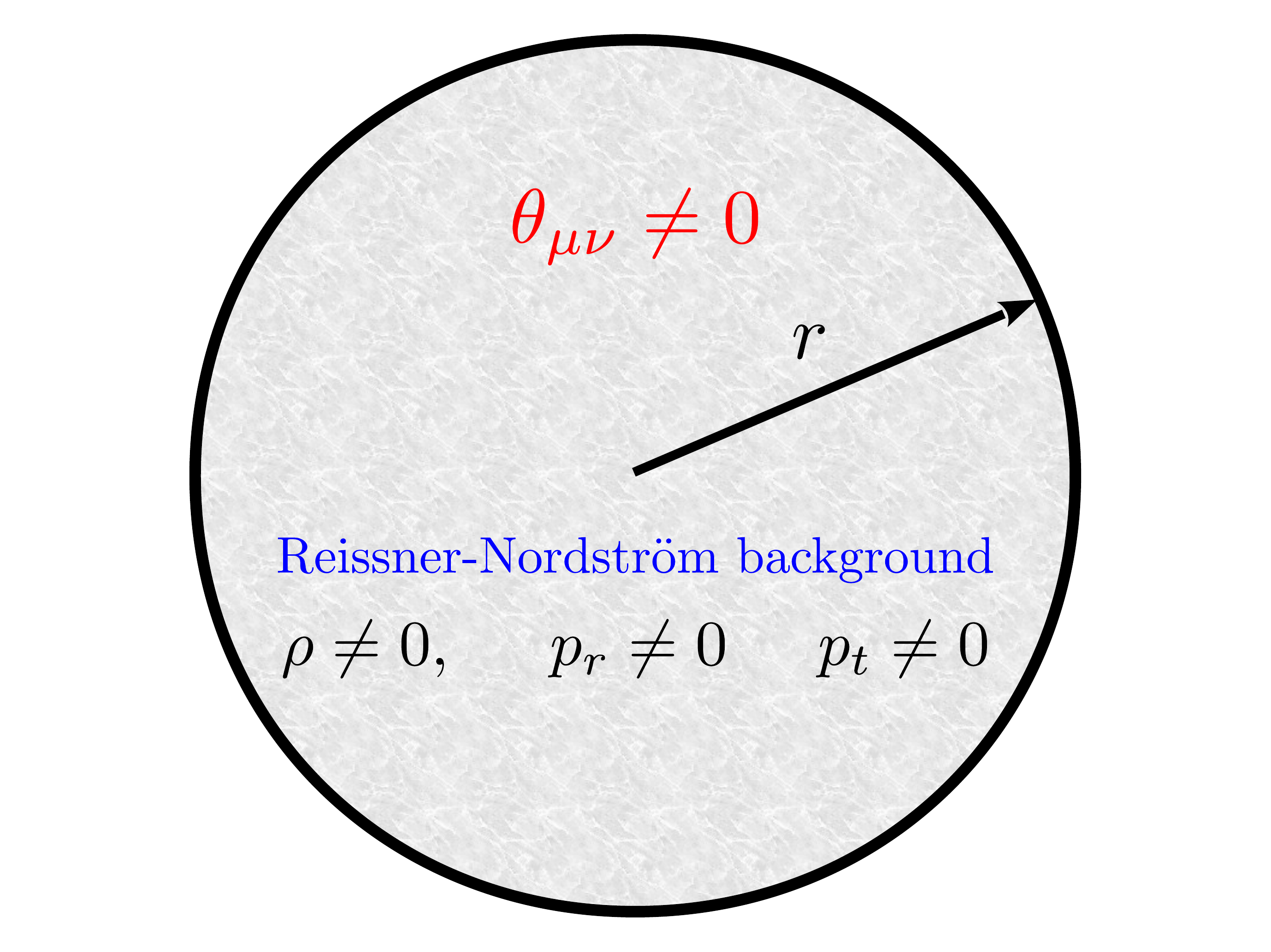}    
\caption{
Spherical symmetric space--time covered with two contributions: i) the Reissner-Nordstr\"om background and ii) the source $\theta_{\mu}^{\nu}$. Of course, the case $\theta_{\mu \nu} \rightarrow 0$ yields the anisotropic black hole solution.
}
\label{fig:1}
\end{figure}
In what follows, we will take a few concrete cases to exemplify the details of this method.

\subsection{Particular constraint $\#$ 1} \label{PCA}

First, let us assume that $\theta_0^0 = \theta_2^2$, which corresponds to $a=0$ and $b=1$.
In this case the deformation function takes the form
\begin{align}
f(r) = \ & \left(\frac{L}{r}\right)^2 \left[1 - \frac{3 M}{r}  +  \frac{2Q^2}{r^2} \right]^{-2} 
e^{\xi(r)},
\end{align}
and therefore the metric function is computed to be
\begin{align}
e^{-\lambda} &= 
\Bigg[
1 + \alpha \left(\frac{L}{r}\right)^2 \left[1 - \frac{3 M}{r}  + \frac{2Q^2}{r^2} \right]^{-2}
\Bigg] 
e^{\xi(r)}.
\end{align}
The components of $\theta_{\mu \nu}$ can be easily computed to obtain
\begin{align}
  \kappa \theta_{0}^{0} &= \left(\frac{L}{r^2}\right)^2 
 \frac{\left(
 \frac{Q^2}{r^2}\left(\frac{5M}{r}-3\right) 
 + 
 \left(1-\frac{M}{r}\right)-\frac{2 Q^4}{r^4}\right)}{\left(-\frac{3 M}{r}+\frac{2 Q^2}{r^2}+1\right)^3},
\\
    \kappa \theta_{1}^{1} &= \left(\frac{L}{r^2}\right)^2 
    \frac{
    \left(\frac{Q}{r}-1\right)\left(\frac{Q}{r}+1\right)}{\left(-\frac{3 M}{r}+\frac{2 Q^2}{r^2}+1\right)^2}.
\end{align}
Moreover, it is easy to verify that the above solution satisfies the condition of energy conservation (\ref{con1d}), as it should be. The fluid parameters can be computed using the equations (\ref{tildero}), (\ref{tildepr}) and (\ref{tildept}) or simply by replace the deformed potential into the Einstein field equations. Thus, the corresponding effective quantities are given by (taking $\kappa=1$) 
\begin{align}
 \tilde{\rho} &= \frac{Q^2}{r^4} + \alpha \theta_0^0 ,
\\
\tilde{p}_r &=  -\frac{Q^2}{r^4} - \alpha \theta_1^1 ,
\\
  \tilde{p}_t &= \frac{Q^2}{r^4} - \alpha \theta_0^0 .
\end{align}
At this level, we can verify by simple inspection that an additional anisotropic term naturally emerges. In particular, in light of the MGD approach, the anisotropy can always be written as
\begin{align}
    \Delta \equiv \Delta_0 + \alpha \Delta_1 ,
\end{align}
where $\Delta_0$ encode the usual RN anisotropy, and $\Delta_1$ is directly linked to the MGD method. So, for this particular example, we have
\begin{align}
    \Delta &= 
    \frac{2 Q^2}{r^4} 
    - 
2 \alpha  
    \left(\frac{L}{r^2}\right)^2    
    \frac{
    \left(1-\frac{2 Q^2}{r^2}\right) 
       }{\left(-\frac{3 M}{r}+\frac{2 Q^2}{r^2}+1\right)^3} 
       \text{e}^{\xi(r)} .
\end{align}
It is crucial to point out that when $\alpha\to0$ the RN anisotropy is recovered. Finally, to check for potential singularities, we compute the Ricci scalar as well as the Kretschmann scalar, which are found to be

\begin{align}
R =  & \ R_0 -2 \alpha  \left(\frac{L}{r^2}\right)^2 \frac{\left(\frac{Q^2}{r^2}\left(\frac{6 M}{r}-5\right) - \frac{2 Q^4}{r^4}+1\right)}{\left(-\frac{3 M}{r}+\frac{2 Q^2}{r^2}+1\right)^3}, 
\label{scp1}\\  
\begin{split}
K \approx &\ K_0 \ +
8 \alpha  \left(\frac{L}{r^3}\right)^2
\Bigg[
\frac{\frac{17 Q^4}{r^4} \left(1-\frac{4 M}{r}\right) + \frac{18 Q^6}{r^6}}{\left(-\frac{3 M}{r}+\frac{2 Q^2}{r^2}+1\right)^3}  \ +
\\
& \frac{\frac{Q^2}{r^2}\left(\frac{84 M^2}{r^2}-\frac{38 M}{r}+3\right) - \frac{4 M}{r}\left(1-\frac{3 M}{r}\right)^2 }{\left(-\frac{3 M}{r}+\frac{2 Q^2}{r^2}+1\right)^3}
\Bigg]\label{ksp1},
\end{split}
\end{align}
where the classical value of the Ricci scalar is precisely $R_0=0$ and $K_0$ is then
\begin{align} \label{KNR}
    K_0 \equiv \frac{8}{r^4}
    \left(\frac{6 M^2}{r^2}-\frac{12 M Q^2}{r^3}+\frac{7 Q^4}{r^4}\right) .
\end{align}
At this level, some comments are in order. Firstly, as the Ricci scalar is zero in the classical case, only a relevant contribution appears when we turn $\alpha$ on. Thus, the MGD approach introduces a non--trivial deviation absent in the classical counterpart.
Second, the Kretschmann scalar is present in the classical solution and becomes more complicated when $\alpha \neq 0$. Although an exact expression is available, we only focus on the first terms in $\alpha$ to verify the impact of the deformation. Additionally, we quickly check that when $\alpha=0$  and $Q \rightarrow 0$ we recover the well--known solutions for the Schwarzschild black hole case. 
Finally, it is worth noticing that two critical points arise when considering $a=0$ and $b=1$, namely,
\begin{eqnarray}
r_{ce}&=&\frac{3}{2}M+
\sqrt{M^{2}-\frac{8}{9}Q^{2}} ,\\
r_{ci}&=&\frac{3}{2}M-
\sqrt{M^{2}-\frac{8}{9}Q^{2}} .
\end{eqnarray}
As can be checked from Eqs. (\ref{scp1}) and (\ref{ksp1}), the Ricci and  Kretschmann scalars blow up at these points which means that these two points correspond to singularities located at $r>0$. Furthermore, it can be shown that $r_{ce}$ results to be greater than the event horizon located at $r_{+}=M+\sqrt{M^{2}-Q^{2}}$, which means that, $r_{ce}$ is a naked singularity. In this sense, the solution obtained here for $a=0$ and $b=1$ must be considered as an exterior solution of a compact star with radius $R>r_{c}$ 

\subsection{Particular constraint $\#$ 2}

Now we will assume the traceless condition for the corresponding anisotropies. The above is a reasonable consideration because of the electromagnetic theory in 3+1 dimensions satisfy the same condition for $F_{\mu \nu}$. 
In term of the general solution, we reproduce the traceless condition when $a=-1$ and $b=-2$. Thus, in term of the $\theta$-components we have:
\begin{align}
2\theta_2^2 = -\theta_0^0 - \theta_1^1,
\end{align}
where the corresponding solution is 
\begin{align}
    f(r) &= 
    -
    \left(\frac{L}{r}\right) 
    \frac{
    \left[\frac{ M_0 - 3 M + 4 r }{ M_0 + 3 M - 4r } \right]^{\frac{3 M}{2 M_0}}
    }{
    \left[-4 +\frac{6 M}{r}-\frac{2 Q^2}{r^2}\right]^{\frac{1}{2}} }
    \text{e}^{\xi(r)} ,
\end{align}
where we have defined the auxiliary parameter as
\begin{align}
M_0 \equiv \sqrt{9 M^2-8 Q^2}.
\end{align}
Please, notice that $M_0$ is a defined positive quantity which means that $9 M^2 \geqslant 8 Q^2$.
Again, our solution is reduced to the uncharged case demanding $Q \rightarrow 0$ which produce:
\begin{align}
\lim_{Q \rightarrow 0} f(r) \equiv \frac{r-2 M}{2 r-3 M} \left(\frac{L}{r}\right). 
\end{align}
In this case, the conformally deformed Schwarzschild exterior is now
\begin{align}
\text{e}^{-\lambda} = 
\Bigg[
1 -
\alpha
    \left(\frac{L}{r}\right) 
    \frac{
    \left[\frac{ M_0 - 3 M + 4 r }{ M_0 + 3 M - 4r } \right]^{\frac{3 M}{2 M_0}}
    }{
    \left[-4 +\frac{6 M}{r}-\frac{2 Q^2}{r^2}\right]^{\frac{1}{2}} }
\Bigg]
\text{e}^{\xi(r)} .
\end{align}
Now, the deformation function allows us to obtain the effective density and pressures. Such inclusion is, however, not necessary due to the complexity of the expressions involved.
In order to check if a new singularity appears, we will show the Ricci as well as the Kretschmann scalars. Surprisingly, the Ricci scalar is identically zero. Thus, the inclusion of an additional anisotropy does not introduce new singularities at this level.
On the other hand, as was reviewed in the analysis of section \ref{PCA}, the Kretschmann scalar is different to zero in the RN solution, and the expression becomes more complicated in the presence of additional anisotropies. In light of this, we will only focus on the first-order term in $\alpha$ (although an exact expression is available). So, the scalars are given by:
\begin{align}
    R & =  R_0 + \alpha R_1,
    \\
    K & \approx K_0 + \alpha K_1,
\end{align}
where, as we previously said, $R_0$ and $R_1$ are zero, $K_0$ is given by Eq. \ref{KNR} and $K_1$ is found to be
\begin{align}
\begin{split}
    K_1 = &  
       \frac{16 \alpha f(r) \text{e}^{-\xi(r)}}{r^4
     \left[2-\frac{3 M}{r}+\frac{Q^2}{r^2}\right]}
  \Bigg[
    \left(17-\frac{33 M}{r}\right)\frac{Q^4}{r^4} \ + 
  \\
& \ \ \ \ \left(\frac{21 M^2}{r^2}-\frac{16 M}{r}+1\right)\frac{2 Q^2}{r^2}
    +  \frac{7 Q^6}{r^6} -
  \\
& \ \ \ \ \left(\frac{6 M^2}{r^2}-\frac{6 M}{r}+1\right)\frac{3 M}{r}
  \Bigg] .
\end{split}
\end{align}
From the above expression, it is clear that the Kretschmann scalar has two extra critical points located at
\begin{eqnarray}
r_{ce}&=&\frac{3M}{4}
+\sqrt{\frac{9 M^{2}}{16}
-\frac{Q^{2}}{2}} ,
\\
r_{ci}&=&\frac{3M}{4}
-\sqrt{\frac{9 M^{2}}{16}
-\frac{Q^{2}}{2}}.
\end{eqnarray}
It is worth noticing that, in contrast to the discussed in the previous section, in this case, the external critical point lies inside the event horizon, namely $r_{ce}<r_{+}$. However, $r_{ce}$ is greater than the Cauchy horizon of the Reissner--Nordstr\"{o}m located at $r_{-}=M-\sqrt{M^{2}-Q^{2}}$ and, as a consequence, the solution could be interpreted as a black hole with a singularity at $r>0$ and an event horizon given by $r_{+}=M-\sqrt{M^{2}+Q^{2}}$. Alternatively, as in the previous section, the solution could be interpreted as an exterior solution of a star with a radius $R>r_{+}$.

\subsection{Particular constraint $\#$ 3}

\begin{figure*}[ht!]
\centering
\includegraphics[width=0.32\textwidth]{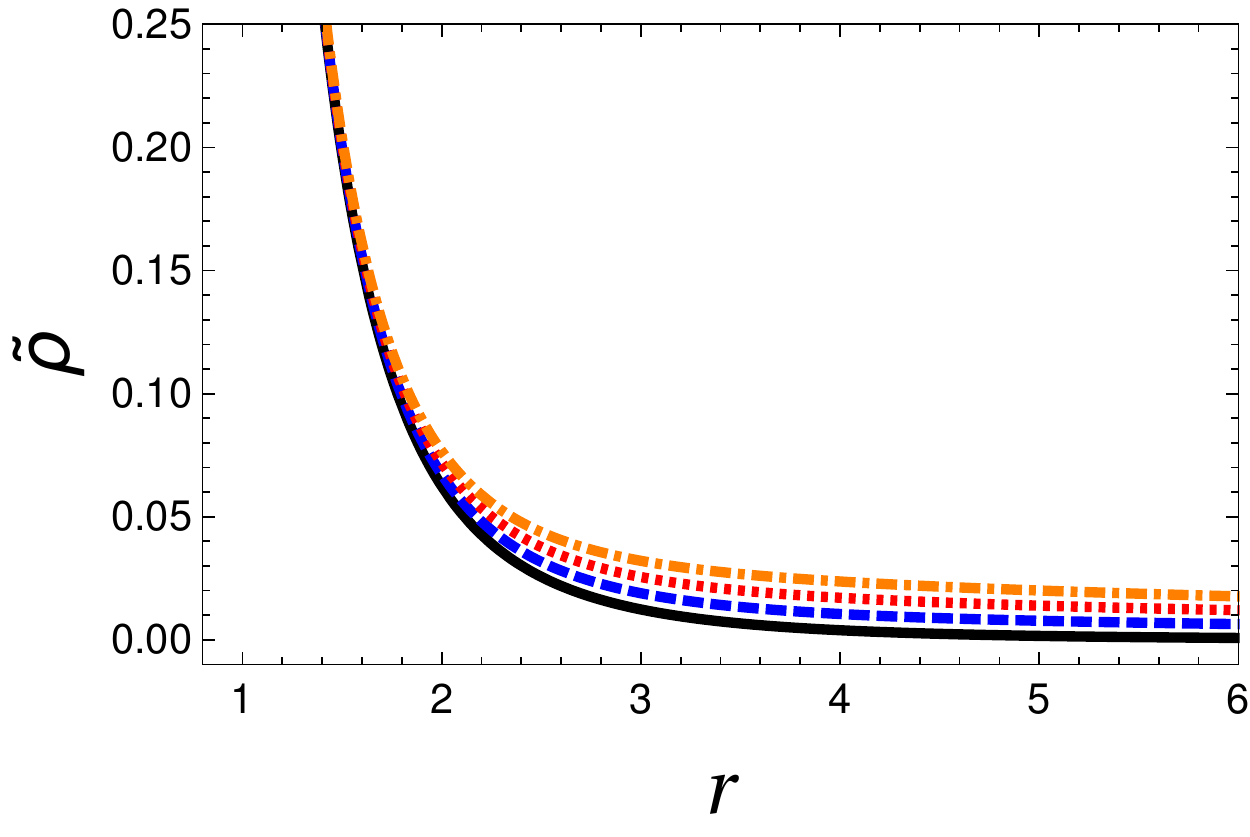}  \
\includegraphics[width=0.32\textwidth]{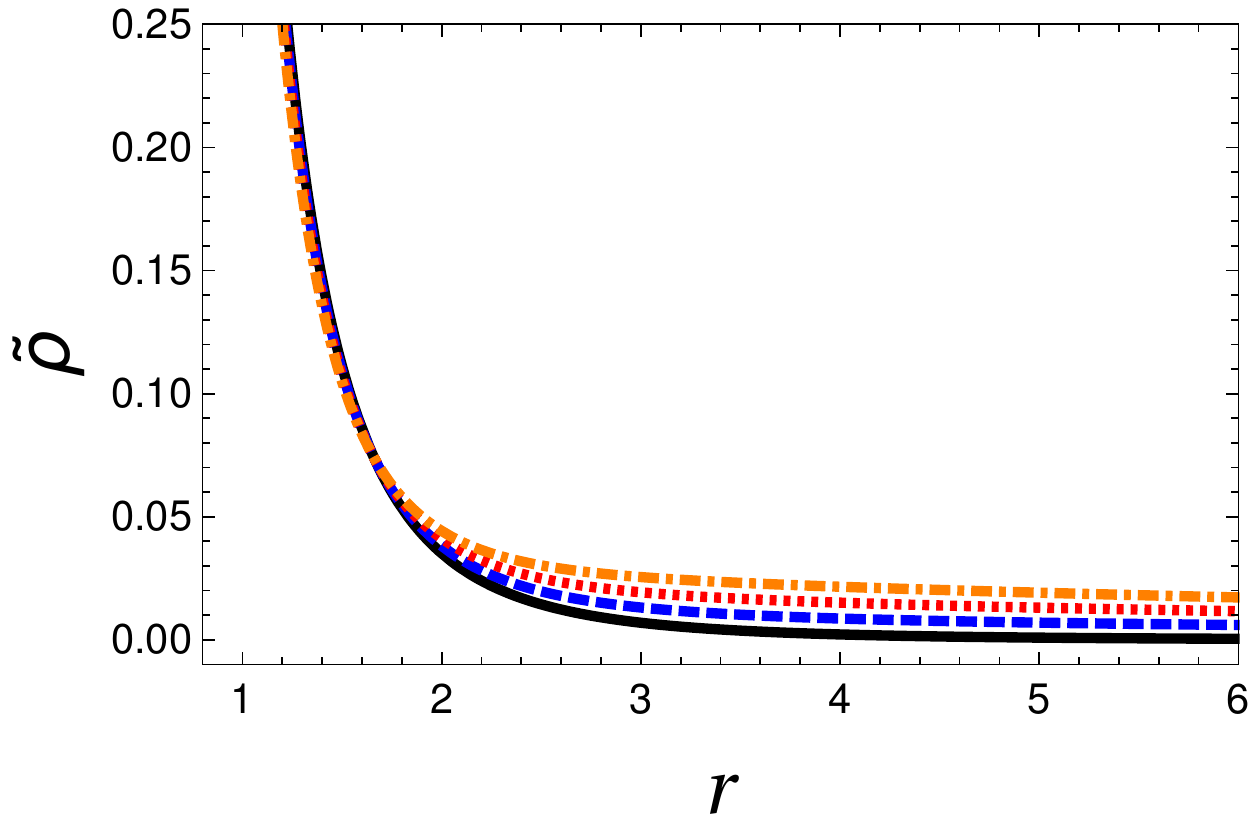}  \
\includegraphics[width=0.32\textwidth]{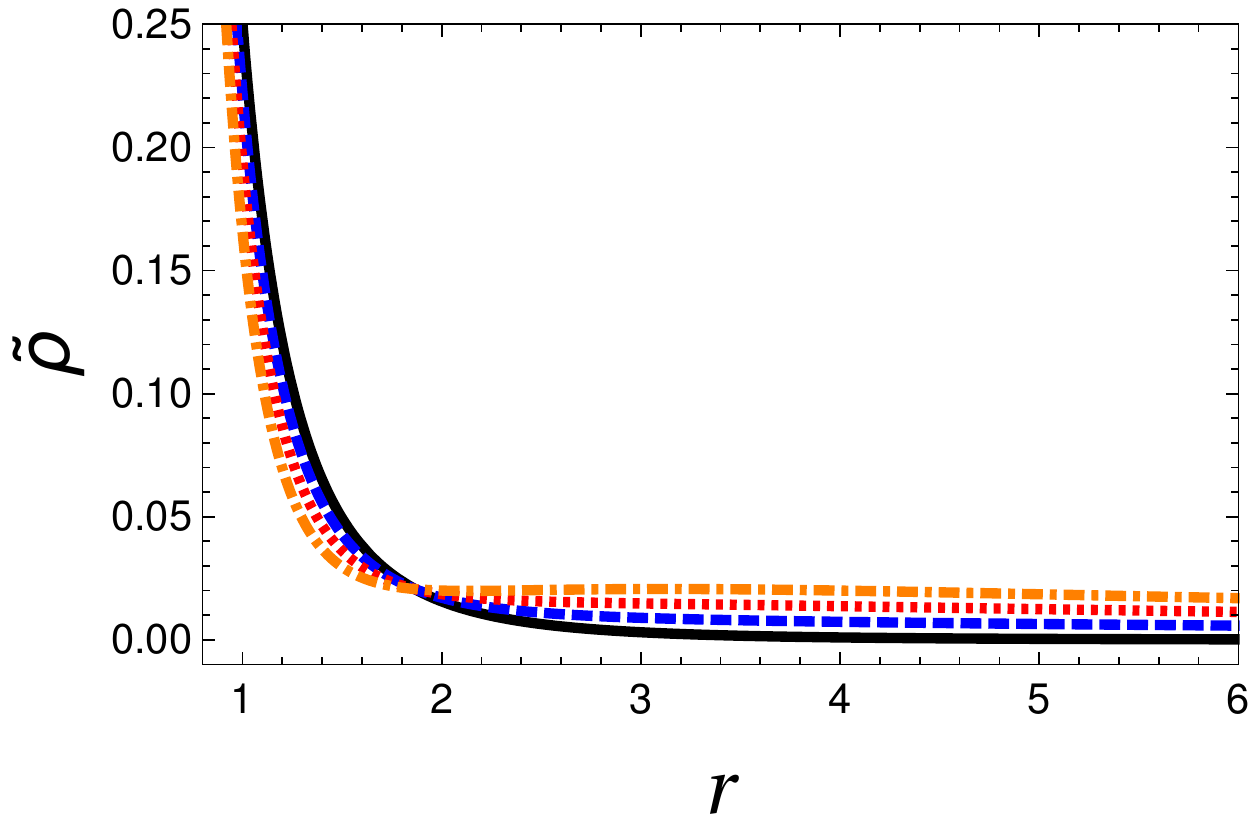}  \

\medskip

\includegraphics[width=0.32\textwidth]{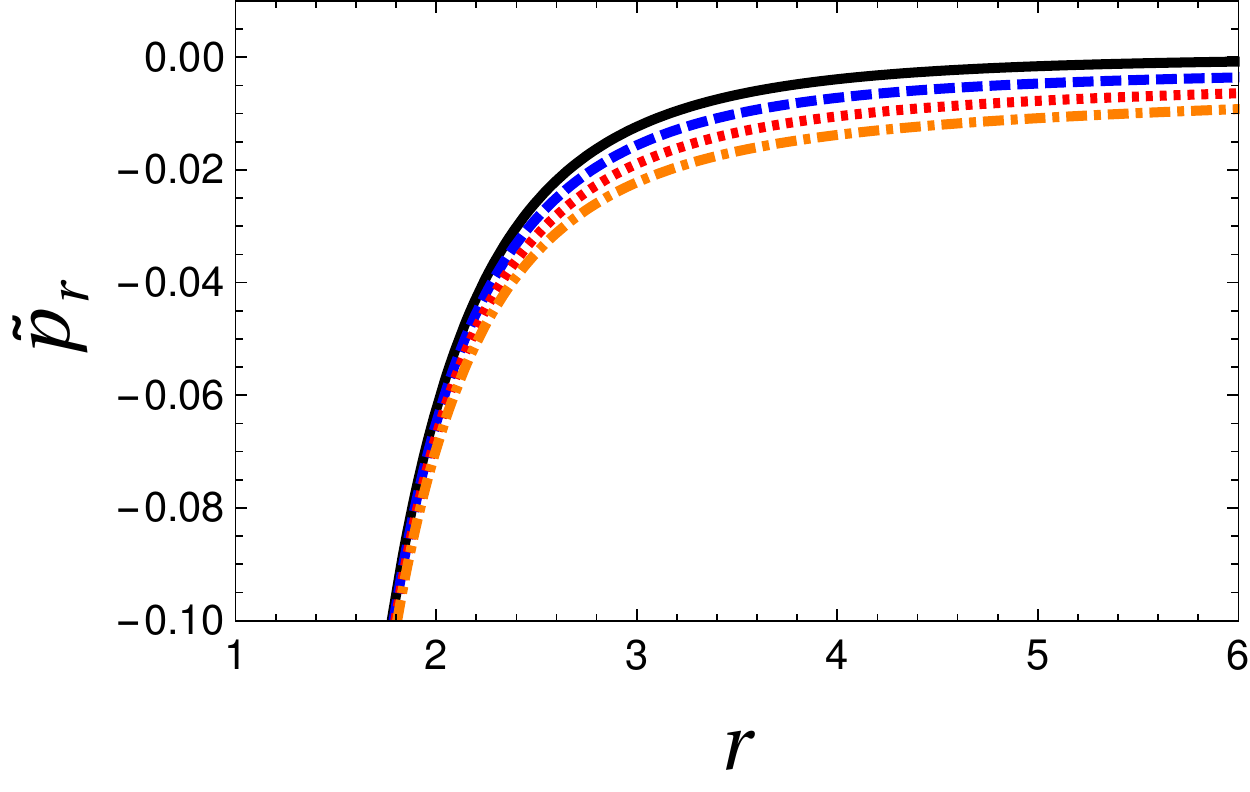} \
\includegraphics[width=0.32\textwidth]{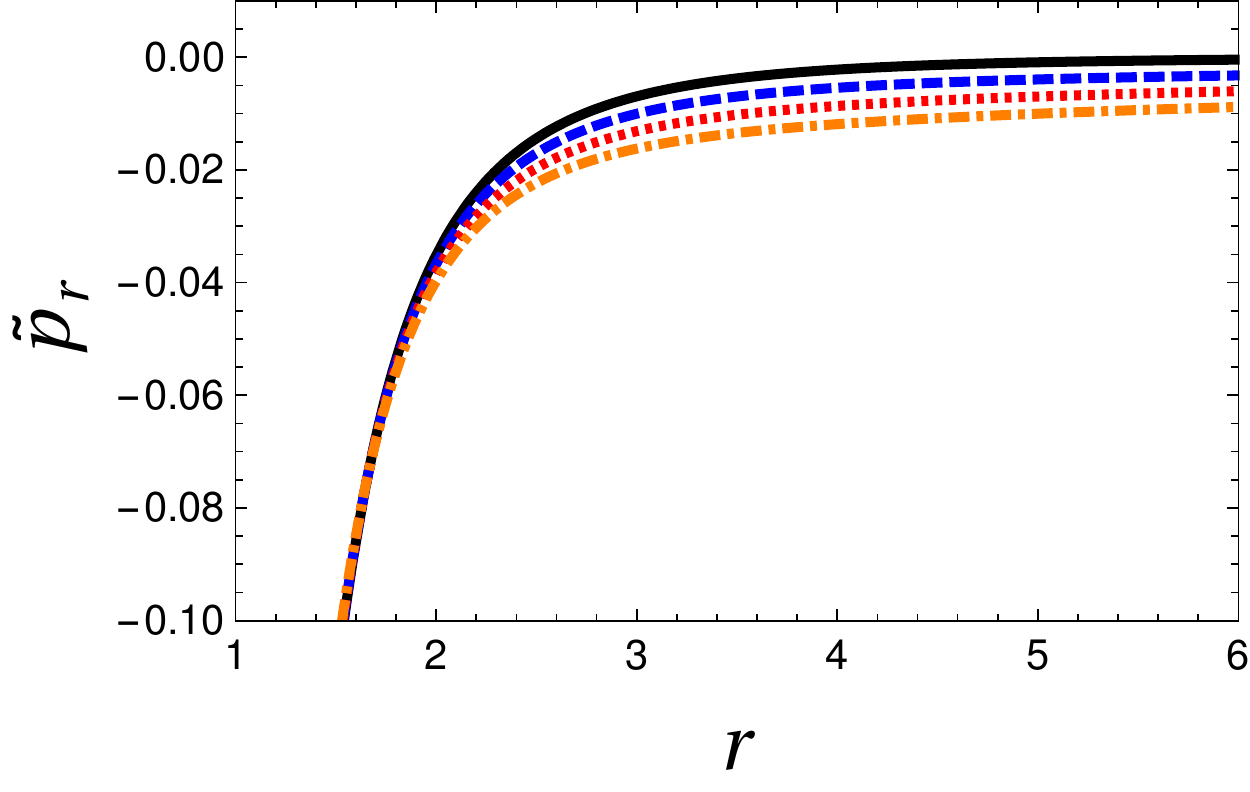} \
\includegraphics[width=0.32\textwidth]{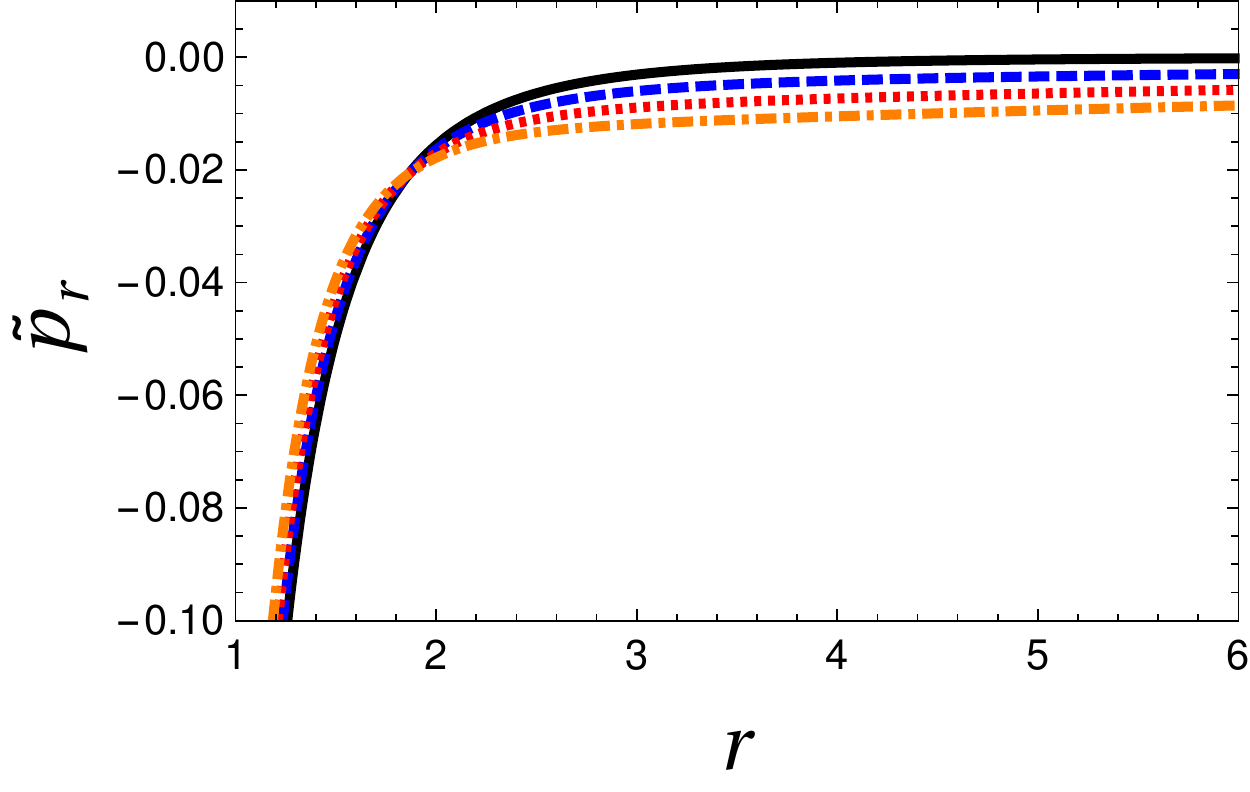}

\medskip

\includegraphics[width=0.32\textwidth]{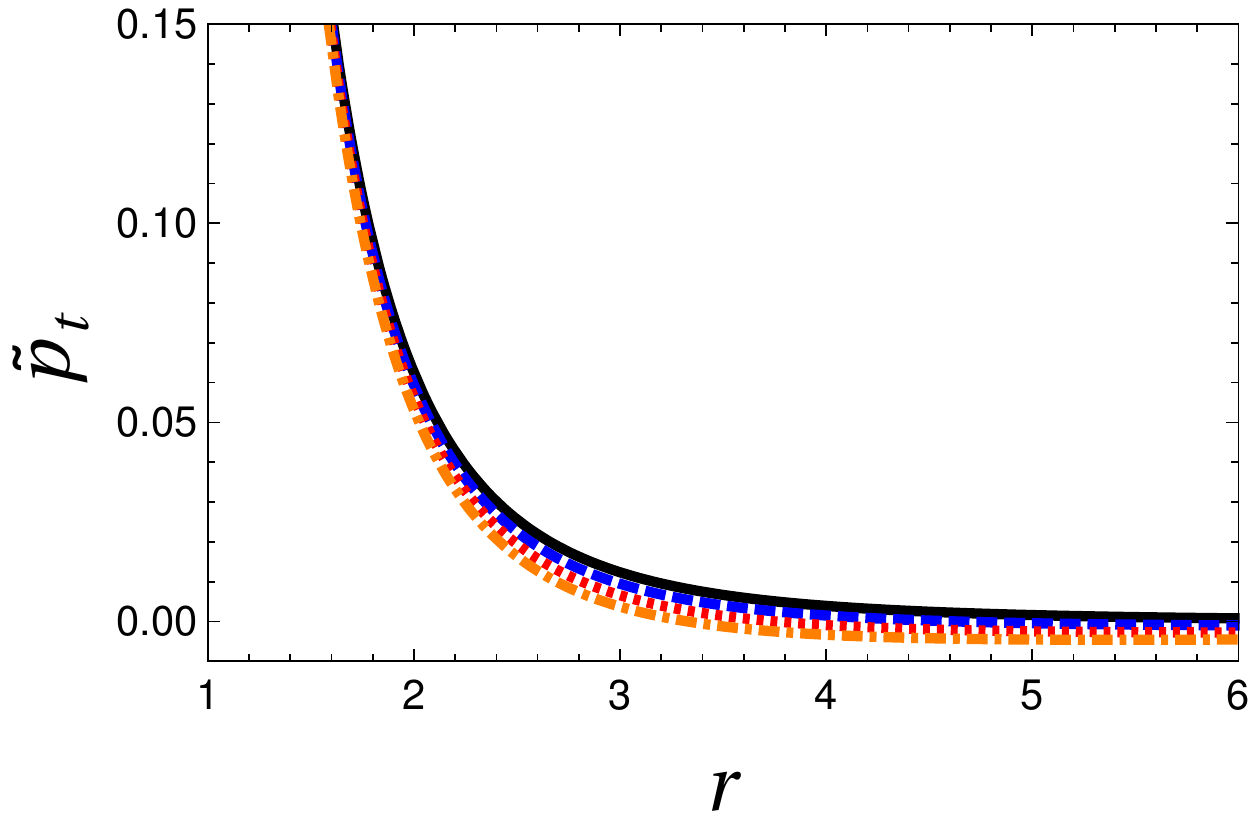} \
\includegraphics[width=0.32\textwidth]{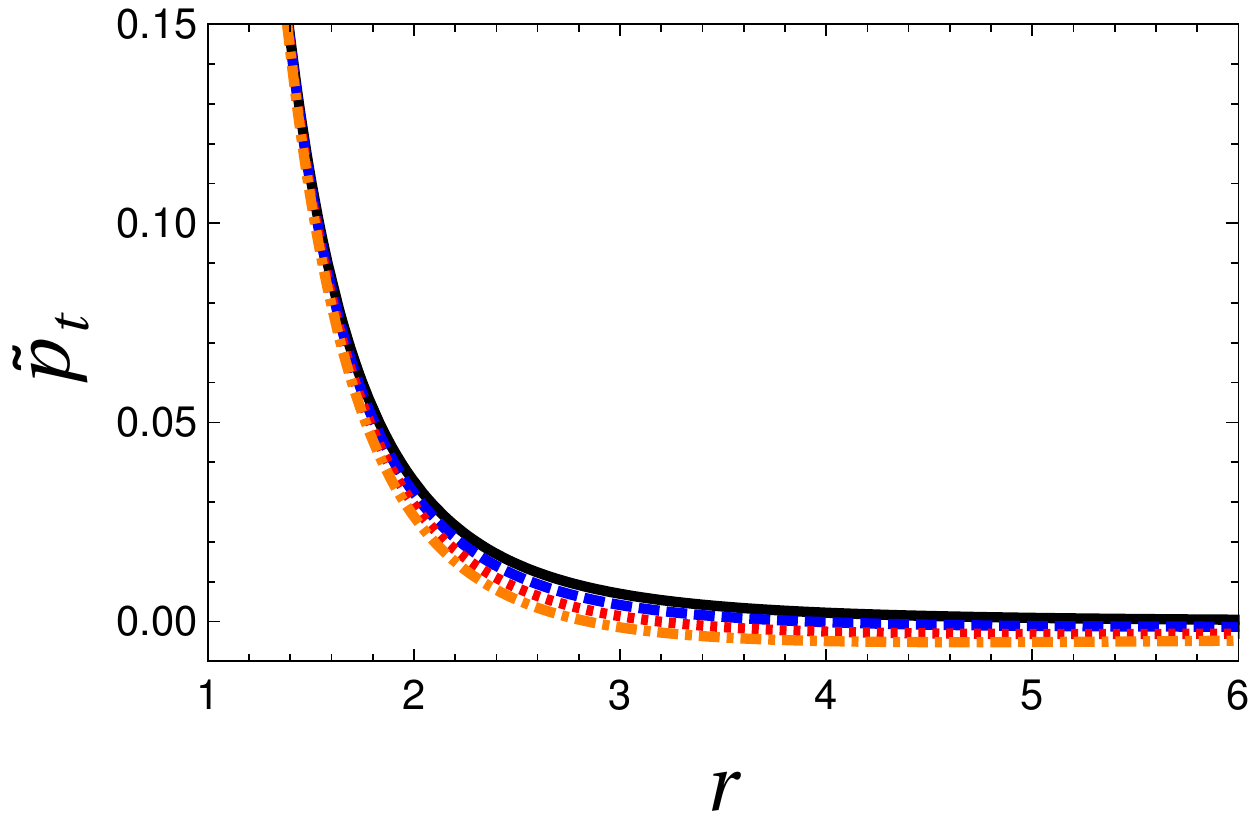} \
\includegraphics[width=0.32\textwidth]{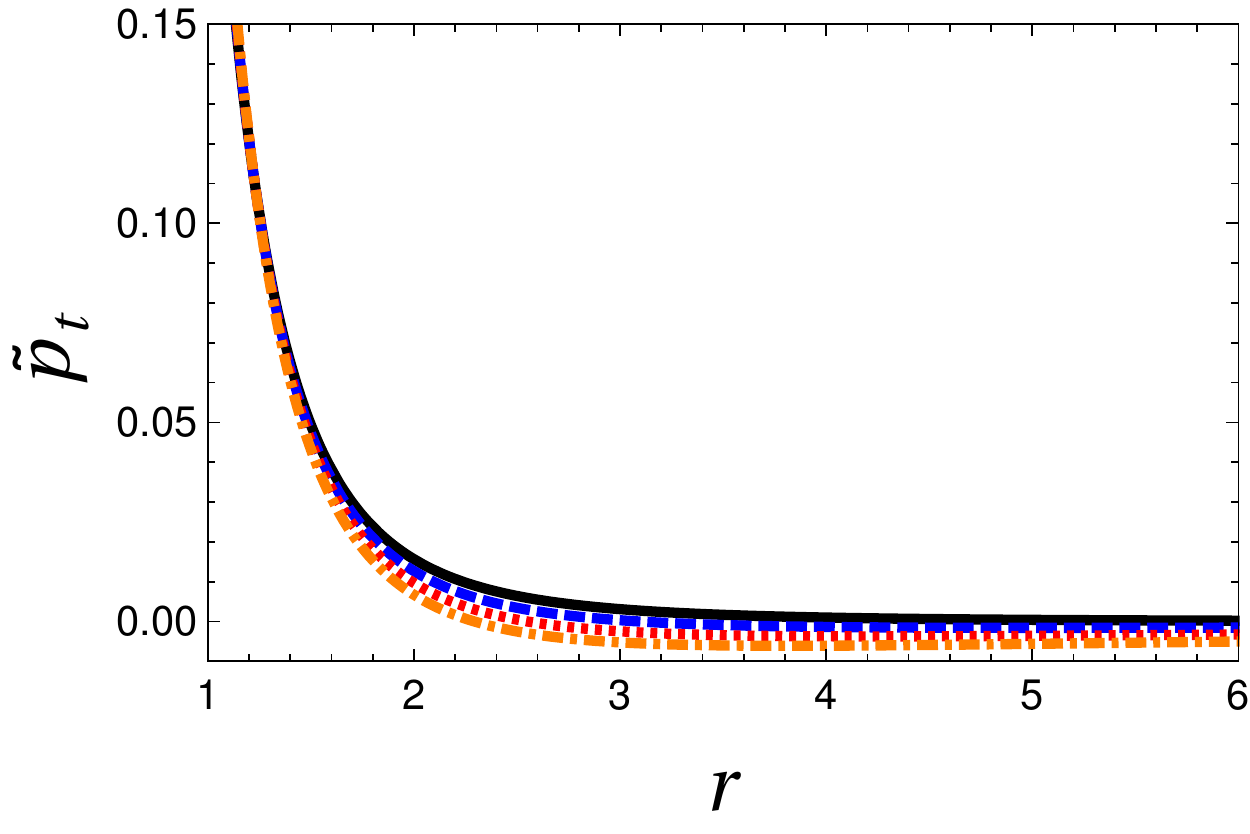}

\caption{\label{fig}
The figures show the evolution of the parameters $\{\tilde{\rho},\tilde{p}_r,\tilde{p}_t\}$ versus the radial coordinate for the third model. We have added the RN case for comparison. To show the impact of the parameter $\alpha$, we evaluate the functions for three different values of the anisotropic coupling $\alpha$ plus the RN solution i.e.: 
i) $\alpha=0$ for the RN solution (solid black line),
ii) $\alpha=-0.025$ (dashed blue line),
iii) $\alpha=-0.050$ (dotted red line) and finally
iv) $\alpha=-0.075$ (dot--dashed orange line).
The first, second and third column correspond to $Q=1$, $Q=0.75$ and $Q=0.5$ respectively. The rest of the parameters are taken to be one. 
}
\end{figure*}

Finally, we will show a new solution without extra singularities. As we previously commented, the crucial point relies on the correct choice of the free parameters $\{a,b\}$. Following the constraint \eqref{b1}, we will take $a=2$ and $b=0$, i.e. 
\begin{align}
\theta^0_0 = 2 \theta_1^1 ,
\end{align}
and solving it, we obtain the corresponding deformation function, which is
\begin{align}
f(r) &= \bigg(\frac{r}{L}\bigg) 
    \left[1 - \frac{2 M}{r} +\frac{Q^2}{r^2}\right]^2 ,
\end{align}
and the deformed metric potential is then given by
\begin{align}
    e^{-\lambda} &=
    \left[ 
    1 + \alpha  
    \bigg(\frac{r}{L}\bigg) 
    \left[1 - \frac{2 M}{r} +\frac{Q^2}{r^2}\right]
    \right]
    e^{\xi(r)}.
\end{align}
As this concrete example is free of singularities, we will compute the complete set of functions, i.e., the thermodynamics functions (density and pressures) as well as the corresponding anisotropies.
The components of the anisotropic tensor are
\begin{align}
    \kappa \theta_0^0 &= \frac{2 \alpha }{L r}\left(\frac{Q}{r}-1\right) \left(\frac{Q}{r}+1\right) \text{e}^{\xi(r)} ,
    \\
    \kappa \theta_1^1 &=  \frac{ \alpha }{L r}\left(\frac{Q}{r}-1\right) \left(\frac{Q}{r}+1\right) \text{e}^{\xi(r)} ,
    \\
\kappa \theta _2^2 &= \frac{\alpha }{2 L r} \left(-\frac{3 M Q^2}{r^3}-\frac{M}{r}+\frac{2 Q^4}{r^4}+\frac{Q^2}{r^2}+1\right) ,
\end{align}
whereas the fluid parameters are (taking $\kappa=1$)
\begin{align}
    \tilde{\rho} &= \frac{Q^2}{r^4} + \alpha \theta_0^0 ,
    \\
    \tilde{p}_r &= -\frac{Q^2}{r^4} - \frac{1}{2}\alpha \theta_0^0 ,
    \\
    \tilde{p}_t &= \frac{Q^2}{r^4} - \alpha \theta_2^2 ,
\end{align}
and as always, the anisotropic term is
\begin{align}
\begin{split}
    \Delta = 
\frac{2 Q^2}{r^4}
+
    \frac{\alpha}{2 L r} 
    \Bigg[ & \frac{Q^2}{r^2}\left(1-\frac{7 M}{r}\right) + 
\\
&    
    \frac{3 M}{r}+\frac{4 Q^4}{r^4}-1
    \Bigg] , 
\end{split}
\end{align}
In figure \ref{fig} the behaviour of the density, $\tilde{\rho}$, the radial pressure, $\tilde{p}_{r}$, and tangential pressure, $\tilde{p}_{t}$, is shown for different values of the MGD parameter, $\alpha$. As can be notice, we only considered negatives values of $\alpha$ in order avoid the appearance of exotic matter content. Indeed, it can be shown that for $\alpha>0$, the density reach negatives values. It is worth noticing that the extra anisotropy induced by the $\theta_{\mu\nu}$ sector, slightly modifies the profiles of the original RN matter sector ($\alpha=0$). We also noticed that when $r \sim r_H$ the deformation introduced by the MGD formalism is practically indistinguishable. Conversely, when $r >> r_H$ the effects of the additional anisotropies are dominant.

\section{Energy conditions}
\begin{figure*}[ht!]
\centering
\includegraphics[width=0.32\textwidth]{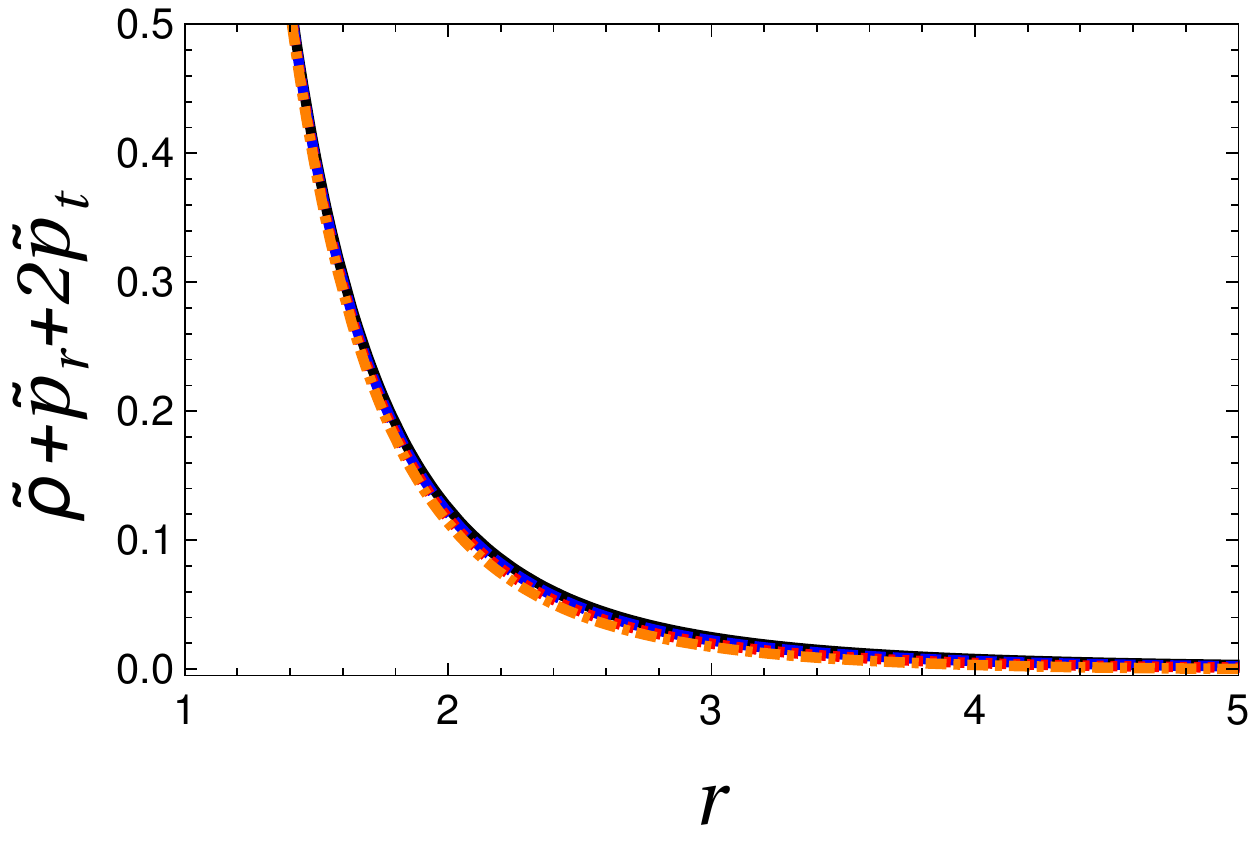} \
%
%
\includegraphics[width=0.32\textwidth]{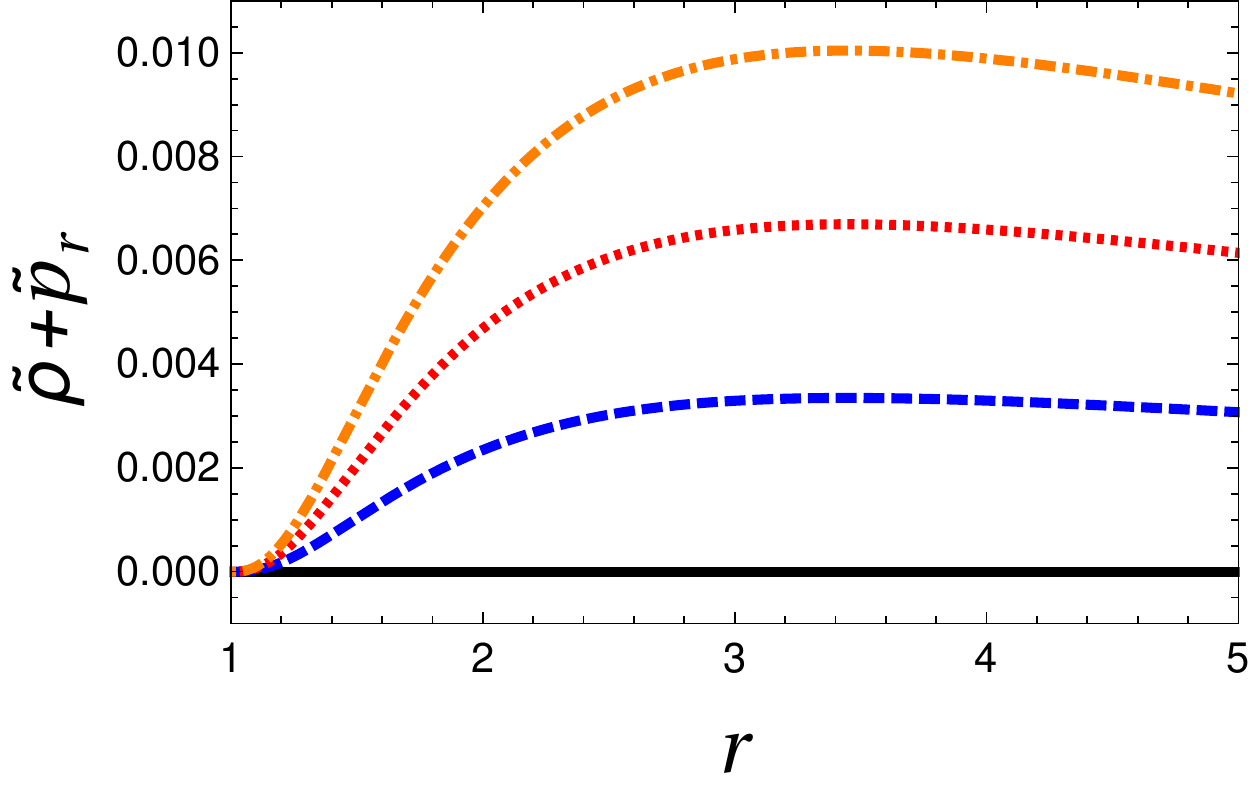}  \
\includegraphics[width=0.32\textwidth]{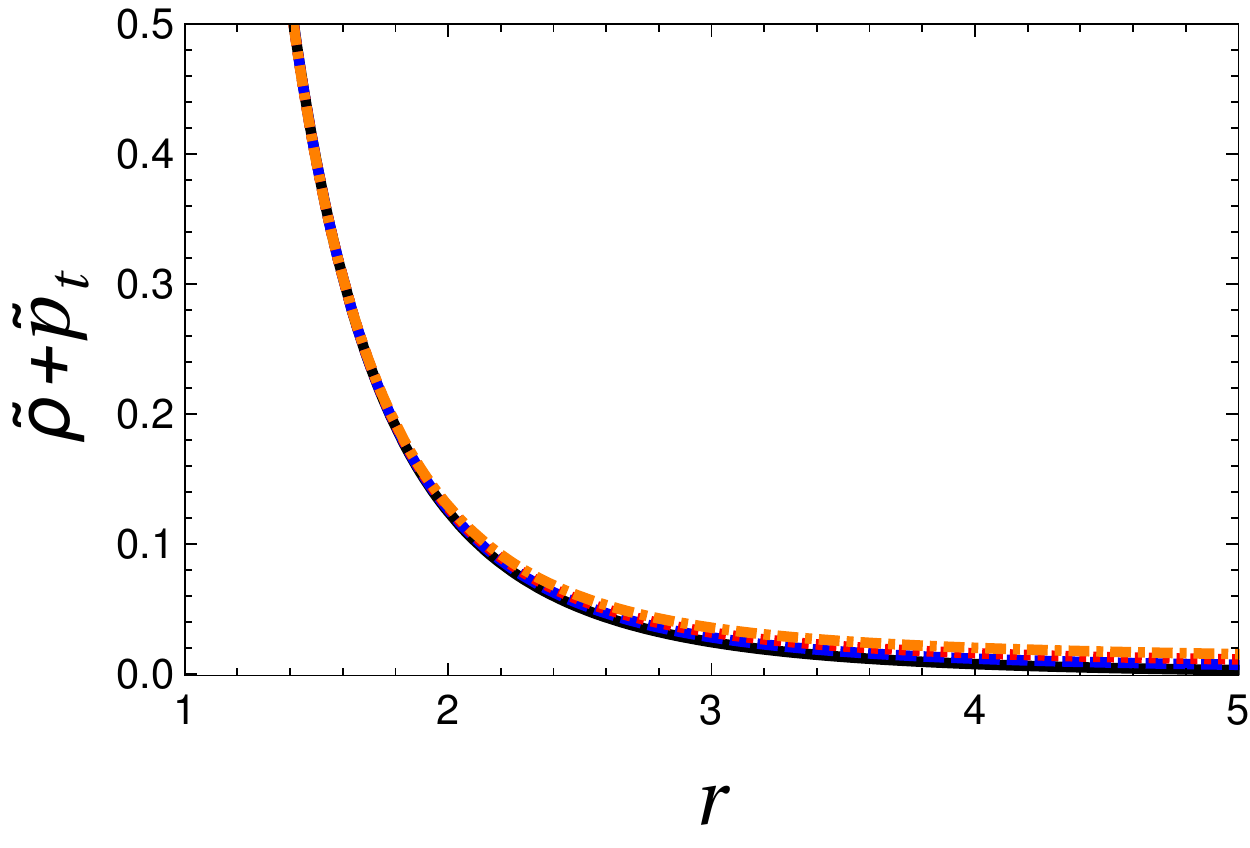}  \

\medskip

\includegraphics[width=0.32\textwidth]{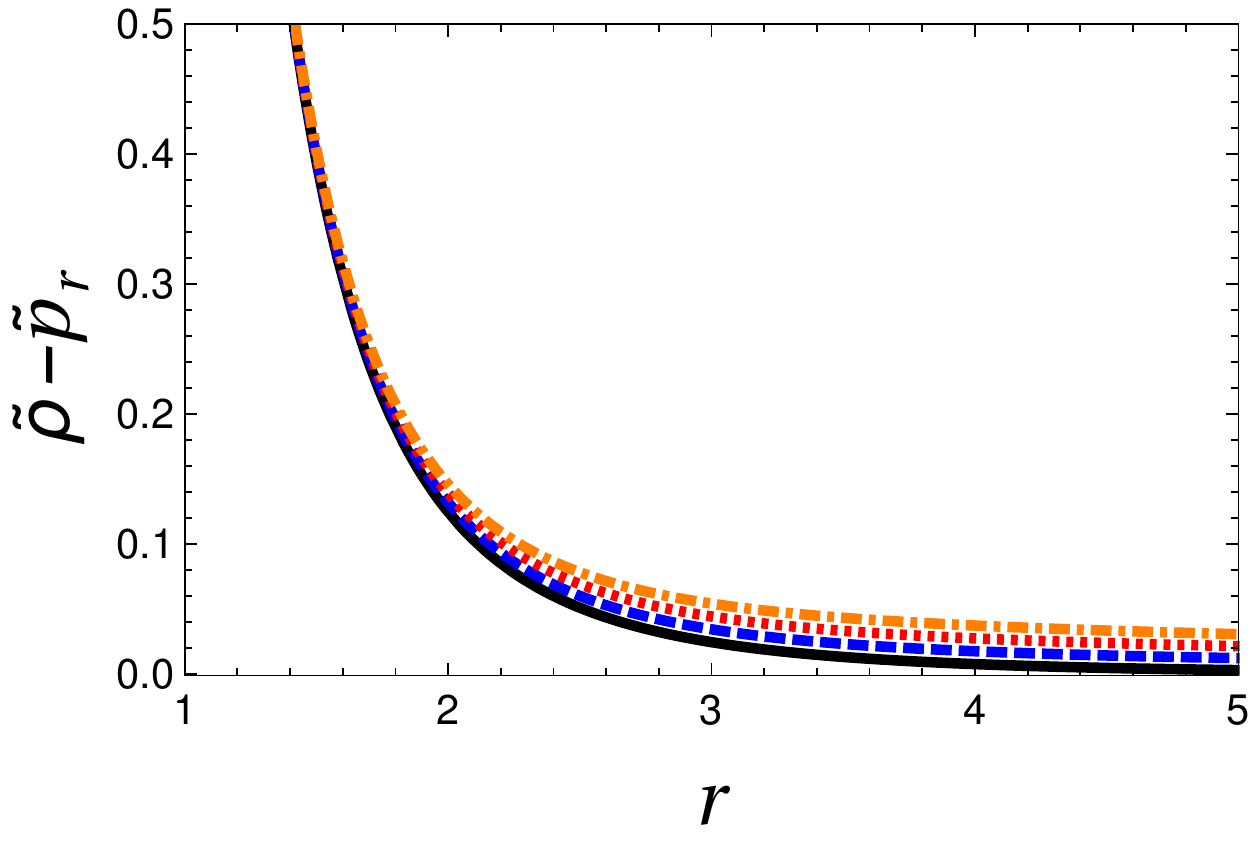} \
\includegraphics[width=0.32\textwidth]{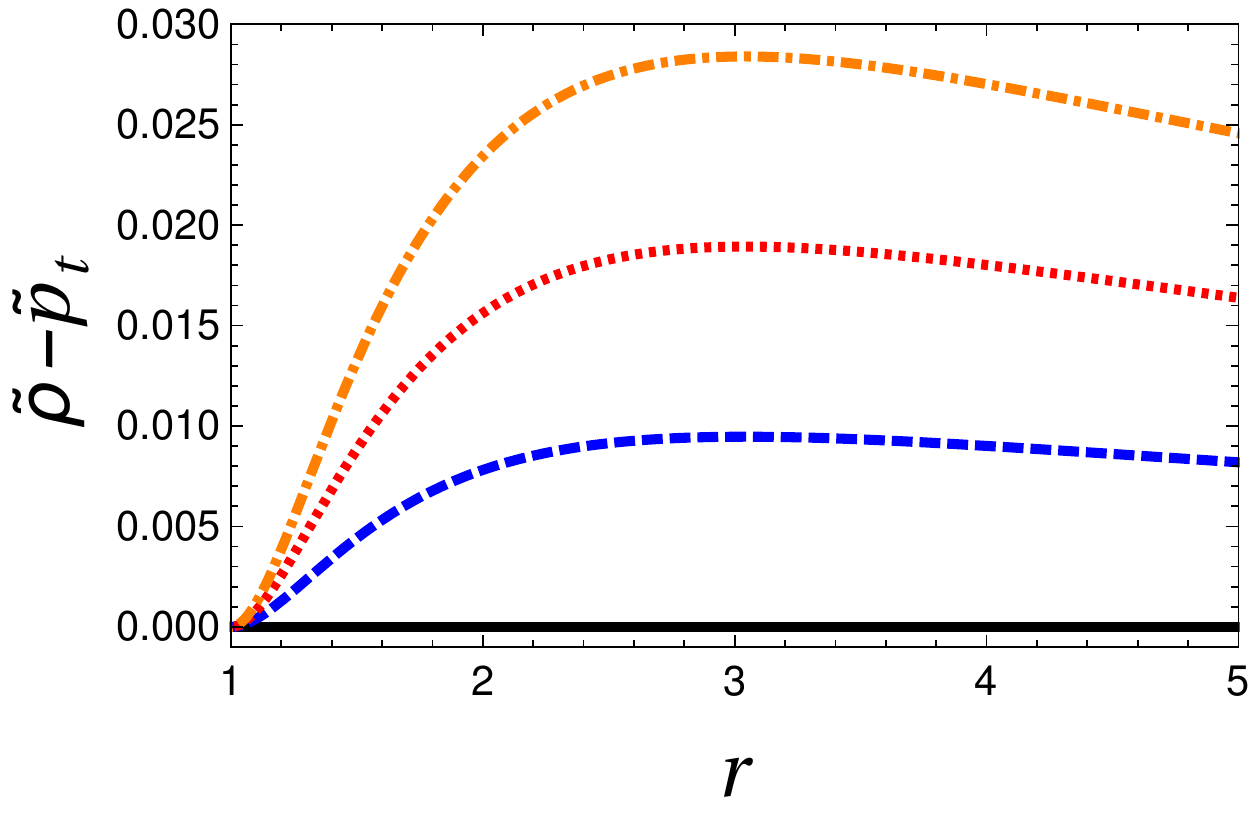} \

\caption{\label{fig3}
The figures show the evolution of energy conditions versus the radial coordinate for the third model. We have added the RN case for comparison. To show the impact of the parameter $\alpha$, we evaluate the functions for three different values of the anisotropic coupling $\alpha$ plus the RN solution i.e.: 
i) $\alpha=0$ for the RN solution (solid black line),
ii) $\alpha=-0.025$ (dashed blue line),
iii) $\alpha=-0.050$ (dotted red line) and finally
iv) $\alpha=-0.075$ (dot--dashed orange line).
The rest of the parameters are taken to be one. 
}
\end{figure*}

This final section is devoted to investigate the corresponding energy conditions for the third model. The energy conditions are usually defined as follow:
\begin{align}
  \text{NEC:} \hspace{2.2cm} \tilde{\rho} & > 0,
    \\
  \text{SEC:} \hspace{0.5cm}     \tilde{\rho} + \tilde{p}_r + 2 \tilde{p}_t & \geq 0,
    \\
  \text{WEC:} \hspace{1.4cm}     \tilde{\rho} + \tilde{p}_r & \geq 0, 
  \hspace{0.5cm}
  \text{and}
  \hspace{0.5cm} 
  \tilde{\rho} + \tilde{p}_t  \geq 0,
    \\
  \text{DEC:} \hspace{1.4cm}     \tilde{\rho} - \tilde{p}_r & \geq 0, 
  \hspace{0.5cm}
  \text{and}
  \hspace{0.5cm}  
  \tilde{\rho} - \tilde{p}_t  \geq 0.
\end{align}
In figure \ref{fig3} we show the energy conditions of the solution including the RN case which corresponds to $\alpha=0$. It is remarkable that, as occurs in the RN solution , all the energy conditions are satisfied for all the values of the MGD--parameter, $\alpha$, here considered. However, we have to mention that the extra anisotropy induce a clear deviation respect to the unperturbed case. To be more presice, the conditions $\tilde{\rho}+\tilde{p}_{r}\ge0$ and $\tilde{\rho}-\tilde{p}_{t}
\ge0$ are saturated by in the RN BH but is strictly possitive in the MGD--deformed solution.

\section{Conclusions}\label{conclu}

To summarize, in the present work we have obtained new exact analytical solutions using the Minimal Geometric Deformation approach on a Reissner-Nordstr\"om background.
Three concrete examples are presented in detail, where the Ricci and Kretschmann scalars are computed too, and the impact of the coupling constant on the solution is investigated. We find that the horizons of the extended solutions coincide with those of the Reissner-Nordstr\"om geometry and the apparition of new horizons or singularities can be avoided by demanding particular constraints on the free parameters appearing in the solutions.
However, it is worth mentioning that the particular equations of states considered in section \ref{RNMGD} could present some extra critical points. Nevertheless, depending on the choice of the parameters involved, this new critical radius may lie inside the Reissner-Nordstr\"om horizons, otherwise, the solution could present naked singularities. Notwithstanding, even in this case, we can still use the cases as valid solutions but considering it as an exterior geometry surrounding a self--gravitating object with a radius greater than any of the critical points.


\section*{Acknowlegements}

The author \'A.~R. acknowledges DI-VRIEA for financial support through Proyecto Postdoctorado 2019 VRIEA-PUCV. F. Tello-Ortiz thanks the financial support by the CONICYT PFCHA/DOCTORADO-NACIONAL/2019-21190856, grant Fondecyt No. 1161192, Chile and projects ANT-1856 and SEM 18-02 at the Universidad de Antofagasta, Chile.


\end{document}